\documentclass[onecolumn,10pt]{revtex4}

\usepackage{graphicx}
\usepackage{dcolumn}
\usepackage{bm}
\usepackage{color}
\usepackage{amssymb,amsmath}

\input epsf
\begin{document}

\title{\Large{On the entropy corrected thermal features of black holes}}

\author{\bf Homa Shababi$^1$\footnote{h.shababi@scu.edu.cn},  Tanwi Bandyopadhyay$^2$\footnote{tanwi.bandyopadhyay@adaniuni.ac.in}, Ujjal
Debnath$^3$\footnote{ujjaldebnath@gmail.com}}

\affiliation{$^1$Center for Theoretical Physics, College of Physical Science and Technology, Sichuan University, Chengdu 610065, People’s Republic of China.\\
$^2$Adani University, Ahmedabad-382421, India.\\
$^3$Indian Institute of Engineering
Science and Technology, Shibpur, Howrah-711 103, India.\\}

\date{\today}

\begin{abstract}
In this work, we investigate the thermal properties of black holes using a new class of generalized entropy functions [K. Ourabah, Class. Quantum Grav., 41, 015010 (2024)]. At the fundamental level, these entropic forms are associated with alternative gravitational laws, within an entropic gravity framework. Our investigation revolves around three distinct entropy functions associated with the Yukawa Potential Correction, Non-local Gravity Correction, and Gradient Field Gravity Correction. Through comparative analysis, we study how such entropic constructs impact the thermodynamic behavior of black holes. For each case, we derive the stability thermodynamic conditions associated with the respective entropic constructs.
\end{abstract}

\maketitle

\section{Introduction}
It is known that the relationship between gravity and thermodynamics has been the subject of extensive research, with pioneering works by Bekenstein, Hawking, and Jacobson \cite{Bekenstein,Hawking,Jacobson} establishing foundational principles in this area. Bekenstein's work on black hole physics and Hawking's contributions on the thermodynamic properties of black holes have been instrumental in uncovering the thermodynamic nature of these celestial bodies. Additionally, Jacobson's groundbreaking finding that the Einstein equation can be cast as an equation of state has provided a novel perspective on the interplay between gravity and thermodynamics. These seminal contributions have paved the way for further investigations into the intricate connections between these two fundamental fields. Recent investigations into the thermodynamic nature of gravity have yielded fascinating insights, with many researchers positing that gravity may be an emergent phenomenon arising from entropic forces (e.g., \cite{Padmanabhan1, Padmanabhan2, Elizalde, Chirco}). Building upon earlier works by Sakharov \cite{Sakharov, Visser} and others, these studies suggest that gravitational forces may be a consequence of changes in informational relationships between material bodies, rather than a fundamental force in its own right. This perspective has opened up new avenues for investigating the underlying mechanisms of gravity and could lead to a deeper understanding of its nature and origins. The concept of holography has captivated scholars in the field of science for numerous decades, and it is gaining increasing prominence across various scientific domains, including the study of black holes (see \cite{Bohm,Talbot,Susskind} and some references therein.)
The connection between entropy and the Bekenstein-Hawking area law has shed new light on the nature of gravity, suggesting the possibility of an emergent phenomenon within this framework. While further research is necessary to confirm this hypothesis, the potential implications of this paradigm are profound and have spurred widespread investigation into its underlying mechanisms and consequences \cite{Cai,Qiu,Komatsu,Shi}. This line of inquiry holds the promise of deeper insights into the fundamental nature of gravity and its relationship to thermodynamic principles. It is important to note that scientific understanding is constantly evolving, and the current consensus on a topic may change as new evidence and research emerges. Hence, it is crucial to stay up to date with recent developments in the field and to critically evaluate the credibility and reliability of sources when seeking the most current and accurate information.
Another developments in the field have revealed deeper connections between gravity and thermodynamics, as exemplified by Verlinde's seminal work \cite{Verlinde}. Building upon this foundation, numerous researchers have explored various generalizations of Verlinde's ideas \cite{Sheykhi1,Meissner,Sheykhi2,Das,Nicolini,Merino,Obreg,Moradpour,Gao,Pazy,Abreu1,Ourabah1,Nozari1,Nozari2}, with a common focus on devising alternative formulations of gravitational interactions that are informed by thermodynamic principles. These investigations often begin with modifications to the Bekenstein-Hawking area law and explore the resulting implications for Newtonian gravity, shedding new light on the intimate relationships between gravity and entropy. In a recent contribution \cite{Kamel}, this direction of inquiry was reversed, with the author seeking to establish a thermodynamic basis for well-known, yet theoretically unjustified, generalizations of Newtonian gravity. Such generalized gravitational laws are widely studied in astrophysics and cosmology \cite{plus1,plus2,plus3,plus4}, but lack a strong theoretical justification. This innovative approach holds the potential to deepen our understanding of the fundamental nature of gravity as it suggest an origin for these alternative gravity models based on a generalized entropic form. The study in \cite{Kamel} investigated various modified versions of Newtonian gravity, categorized into two primary classes: those affecting long-range interactions (comparable to or greater than the scale of the solar system) \cite{Maneff1,Maneff2,Maneff3,Hagihara,Tohline,Kuhn,Acedo,Finzi,Giusti} and those affecting short-range interactions (typically below the millimeter scale) \cite{Hoyle,Smullin,Modesto1,Biswas,Modesto,Lazar}. The modifications in the first category aim to address the observed flat rotation curves of galaxies without invoking dark matter, while those in the second category seek to resolve the singularity of the Newtonian potential at short distances by weakening the strength of the Newtonian interaction at close proximity. These diverse approaches highlight the ongoing efforts to understand the behavior of gravitational systems across different scales and to develop theoretical frameworks that can account for observed phenomena in a more comprehensive manner. For the second category, which we are more interested in our work, the author of Ref. \cite{Kamel}, focused on small-scale deviations from Newtonian gravity. This included an examination of modified Newtonian gravity in the form of a Yukawa-type potential, as well as nonlocal gravity of exponential type and gradient modifications \cite{Hoyle,Smullin,Modesto1,Biswas,Modesto,Lazar}. One of the main results of \cite{Kamel} was the author investigated the relationship between various gravitational laws and their entropic forms in the context of entropic gravity.
In what follows, we explores the thermodynamic behavior of modified gravity theories that deviate from Newtonian gravity at small scales. We focus on three specific models: Yukawa-type potential, nonlocal exponential gravity, and gradient modifications. We begin by examining the equipartition law for each model and then proceed to calculate the heat capacity, in order to assess the thermodynamic stability of the modifications. Our aim is to determine whether these models exhibit a consistent and coherent thermal behavior, which could provide insight into their viability as potential alternatives to classical gravity.

\section{Corrections in Entropy Functions}

 Modifications in the entropy function are being incorporated in various forms to attune with the emerging physics of string theory and loop quantum gravity as well as from the phenomenological perspective \cite{Meissner,Kaul,Adler,Das1,Myung,Doma,Chatterjee,Ghosh,Camelia,Nozari,Barun,Barun2}. These studies have indicated in general, the presence of a logarithmic correction term in the formulation of the entropy function. However, in our current work, we focus on analyze three recent modified forms \cite{Kamel} of the entropy function. These generalized entropies are associated with various alternative gravity laws, widely used in the literature, through the Verlinde's entropic gravity approach. Motivated by the work \cite{Abreu}, we study the thermal characteristics for these three different entropy formalism concerning to the black hole horizon. These corrections are ascribed to the deviations from the Newtonian gravity in the near field. Throughout the work, we consider $\hbar=c=k_B=1$.

\subsection{Yukawa Potential Correction}

The first small-scale deviation from the standard Newtonian gravitational potential is caused by the Yukawa potential $\phi=-\frac{GM}{R}(1+\alpha e^{-\frac{R}{\lambda}})$, where the two parameters $\alpha$ and $\lambda$ are respectively the strength and range of the corrected term. The Newtonian potential is retrieved for $\alpha=0$. Extensive studies on this potential function from both theoretical and experimental aspects have led to certain constraints on the parameters. Due to this changed potential, the entropy function gets modified to the form \cite{Kamel}

\begin {equation}\label{YukEntOrig}
S_{Y}=\frac{1}{4G}\left[A-2\alpha{\lambda}^2 e^{-\sqrt{\frac{A}{4\pi{\lambda}^2}}}\left(12\pi+6\sqrt{\frac{\pi A}{{\lambda}^2}}+\frac{A}{{\lambda}^2} \right) \right]
\end{equation}

Here $G$ is the gravitation constant and $A=16\pi G^2 M^2$ is the horizon area of a Schwarzschild black hole with source mass $M$. The above equation reduces to the standard Bekenstein-Hawking area law for $\alpha=0$. Up to now, some efforts have been done to obtain the values of $\alpha$ and $\lambda$. In experimental investigations, Kapner et al. explored the gravitational inverse-square law at scales below the dark-energy length \cite{Kapner}. They discovered that, for a strength parameter $|\alpha| = 1$, the characteristic length should be less than $56 \mu m$. Additionally, a recent experiment conducted by Lee et al. \cite{Lee} revealed an even smaller value for $\lambda$, i.e. $\lambda < 38.6 \mu m$. However, solar system tests indicate that, for a range of $\lambda = 5000 AU$, the strength $\alpha$ must fall within the interval $[2.70-6.70]\times 10^{-9}$  \cite{Martino}, which in an entropic framework, these bounds can impose constraints on the deviation relative to the Bekenstein-Hawking area law.
In  other words, when considering the small-scale effort, it is essential to work with a length scale $\lambda$ below the micrometer level, which implies that $\lambda\ll A$. On the other hand, in cosmological contexts, the range of $\lambda$ is significantly larger. However, the correction is dampened by an exceedingly small value of $\alpha$ which is typically around $10^{-9}$ \cite{Martino}.

\begin{figure}
\centering
\includegraphics[width=11cm]{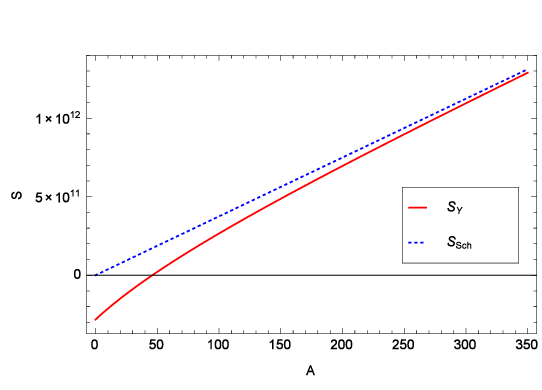}
\caption{\label{fig1}Comparison between Yukawa entropy, $S_Y$, and Schwarzschild entropy, $S_{Sch}$,
versus area $A$ for
$\alpha=1$ and $\lambda=10^{-6}$.}
\end{figure}

Now, to investigate the results, we plotted the Yukawa entropy ($S_Y$), and its Schwarzschild counterpart ($S_{Sch}$) versus horizon area $A$ in Fig.~\ref{fig1}. As it is shown in the figure, when the horizon area of the black hole increases, both Yukawa and Schwarzschild entropies increase. Also, for fixed $A$, it is concluded that $S_{Y}<S_{Sch}$ and this difference decreases with increasing $A$.

Equation \eqref{YukEntOrig} can be rewritten in the following way

\begin{equation}\label{YukEnt}
S_{Y}=\frac{1}{4G}\left[16\pi G^2 M^2-2\alpha{\lambda}^2 e^{-\frac{2GM}{\lambda}} \left(12\pi+\frac{24\pi GM}{\lambda}+\frac{16\pi G^2 M^2}{{\lambda}^2} \right) \right]
\end{equation}

The black hole horizon temperature $T$ is defined as \cite{Abreu}

\begin{equation}
\frac{1}{T}=\frac{\partial S}{\partial M}
\end{equation}

Here $S$ is the specific entropy describing the horizon. Using the corrected entropy form of $S$ from equation \eqref{YukEnt}, the (inverse) temperature becomes

\begin{equation}\label{partial}
\frac{1}{T}=8\pi GM\left[1+\alpha e^{-\frac{2GM}{\lambda}} \left(1+\frac{2GM}{\lambda}\right)\right]
\end{equation}

As we are interested to study the thermal characteristics of the black hole horizon pertaining to the changed form of the entropy function, we would like to cite the connection to the standard equipartition law $M=\frac{1}{2}NT$ as proposed in \cite{Paddy}. The associated number of degrees of freedom (DoF) of the horizon is given by $N=4S$ \cite{Kom}. We shall adhere to the procedure followed in \cite{Abreu}. We shall keep the same number of the DoF based on the Bekenstein-Hawking entropy so that distinct equipartition laws can emerge for specific forms of the horizon entropy.\\

Using \eqref{YukEnt} and \eqref{partial} in the definition of the number of the DoF, we get

\begin{equation}\label{DOF}
\frac{N}{4}=\frac{M}{2T}\frac{\left[1-\alpha e^{-\frac{2GM}{\lambda}} \left(2+\frac{3\lambda}{GM}+\frac{3{\lambda}^2}{2G^2 M^2}\right)\right]}{\left[1+\alpha e^{-\frac{2GM}{\lambda}} \left(1+\frac{2GM}{\lambda}\right)\right]}
\end{equation}

This can be reorganized to produce the new equipartition law, attached to the entropic form (\ref{YukEntOrig}) which is specific to the small-scale Yukawa type correction as

\begin{equation}
M=\frac{1}{2}\left[\frac{1+\alpha e^{-\frac{2GM}{\lambda}} \left(1+\frac{2GM}{\lambda}\right)}{1-\alpha e^{-\frac{2GM}{\lambda}} \left(2+\frac{3\lambda}{GM}+\frac{3{\lambda}^2}{2G^2 M^2}\right)} \right]NT
\end{equation}

The above equation is a new type of generalized equipartition theorem associated with the Yukawa type correction in the Newtonian potential function (see \cite{eqt1,eqt2,eqt3} for generalized equipartition theorem orginated from corrected entropies). It can be considered as equivalent to the standard equipartition law $M=\frac{1}{2}NT$ when the parameter $\alpha$  for this type of correction is taken as zero.\\

\begin{figure}
\centering
\includegraphics[width=11cm]{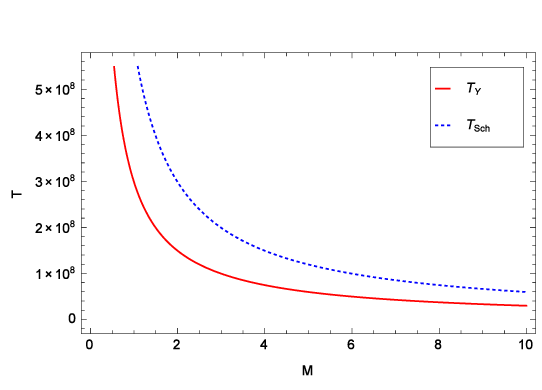}
\caption{\label{fig2}Comparison between Yukawa temperature, $T_Y$, and the Schwarzschild black hole temperature, $T_{Sch}$, versus
black hole mass $M$ for
$\alpha=1$ and $\lambda=10^{-6}$.}
\end{figure}

Then, in Fig.~\ref{fig2} for more comparison, we have depicted the Black hole temperature due to Yukawa corrections, $T_Y$, (Eq.~\eqref{partial}), and the Schwarzschild black hole temperature ($T_{Sch}$,) versus black hole mass $M$. As it is shown, for both models, if $M{1}>M{2}$, we have $T_{1}<T_{2}$. Also, for fixed mass it leads to $T_{Y}<T_{Sch}$.

The heat capacity can be expressed in terms of the entropy function in the following way \cite{Abreu}

\begin{equation}\label{heat}
C=-\frac{[S'(M)]^2}{S''(M)}
\end{equation}
where the operation $'$ implies derivative with respect to $M$. Here the negative sign in $C$ indicates thermodynamic instability of the black hole, as expected. Substituting the modified Yukawa type corrected entropy \eqref{YukEnt} in equation \eqref{heat}, we get

\begin{equation}\label{stability1}
C=-8\pi G M^2 \frac{\left[1+\alpha e^{-\frac{2GM}{\lambda}} \left(1+\frac{2GM}{\lambda}\right)\right]^2}{\left[1+\alpha e^{-\frac{2GM}{\lambda}} \left(1+\frac{2GM}{\lambda}-\frac{4G^2 M^2}{{\lambda}^2}\right)\right]}
\end{equation}

From the above equation, a stable thermodynamic scenario ($C>0$) can be achieved if

\begin{equation}\label{stability}
1+\alpha e^{-\frac{2GM}{\lambda}} \left(1+\frac{2GM}{\lambda}-\frac{4G^2 M^2}{{\lambda}^2}\right)<0
\end{equation}

It is clear from equation \eqref{stability} that the stability of the black holes can never be achieved in a far field (large $\lambda$) with very small strength $\alpha$. This is as per the expected result that a black hole is thermally unstable in nature.

\begin{figure}
\centering
\includegraphics[width=11cm]{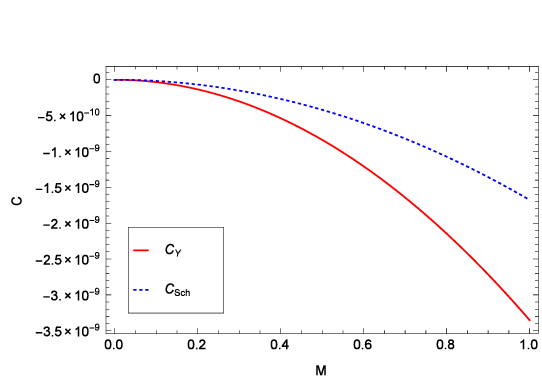}
\caption{\label{fig7}Comparison between Yukawa heat capacity, $C_Y$, and the Schwarzschild heat capacity, $C_{Sch}$, versus
black hole mass $M$ for $\alpha=1$ and $\lambda=10^{-6}$.}
\end{figure}

Afterwards, in Fig.~\ref{fig7}, the behaviors of Yukawa heat capacity $C_Y$ (Eq.~\eqref{stability1}), and the Schwarzschild one $C_{Sch}$, versus $M$ are investigated. According to the plot, for both, we find a negative heat capacity as expected. In other words, as $M$ increases, both heat capacities decrease, but they do so at different rates. Also, for fixed mass, $C_{Sch}>C_{Y}$ and this difference bacome grater with increasing $M$.

\subsection{Non-local Gravity Correction}

Another small-scale variation from the Newtonian gravity surfaced as a result of considering the non-local gravity of the exponential type. The potential function in this case is \cite{nonlocal1,nonlocal2,nonlocal3}

\begin{equation}
\phi=-\frac{GM}{R}~\text{erf}\left(\frac{R}{2l}\right)
\end{equation}

In the theory of the non-local gravity, $l$ is called the characteristic length scale parameter and the function erf $(x)=\frac{2}{\sqrt{\pi}}\int_{0}^{x}e^{-t^2}dt$, is called the Gauss error function with $\lim_{x \to \infty}$erf$(x)=1$. For large distances as compared to the length scale $l$, the given potential reduces to the Newtonian potential. This changed form of the potential generates an alteration to the standard entropy formula as expressed in the following form \cite{Kamel}

\begin{equation}\label{NonLocEnt}
S_N=\frac{1}{4G}\left[12l\sqrt{A}e^{-\frac{A}{16\pi l^2}}+(A-24\pi l^2)~\text{erf}\left(\sqrt{\frac{A}{16\pi l^2}}\right)\right]
\end{equation}

It can be noted from the above expression that the standard area law is attained for large $A$ as compared to the length scale parameter $l^2$. In fact, in that case, the first term inside the bracket simply vanishes and the second term gives only $A$ as the error function approximates to unity.

\begin{figure}
\centering
\includegraphics[width=11cm]{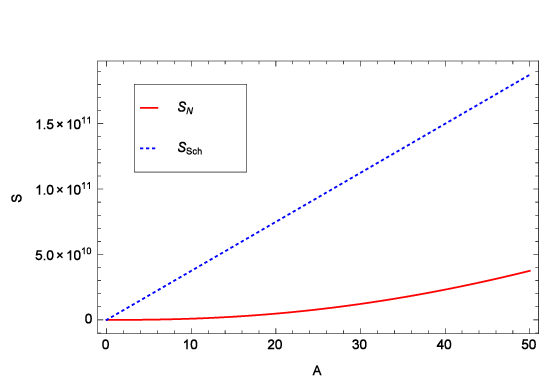}
\caption{\label{fig3}Comparison between Non-local gravity entropy, $S_N$, and Schwarzschild entropy, $S_{Sch}$,
versus area $A$ for $l=10^{-6}$.}
\end{figure}

For this case, as far as we know, for the characteristic length scale parameter $l$, no experimental bounds have been obtained until now, but possible tests, including gravity analogs, may be accessible in the laboratory experiments \cite{moi1}.\\
Now, to present our results, we depicted the modified entropy due to Non-local gravity ($S_N$), and the Schwarzschild entropy ($S_{Sch}$) versus horizon area $A$ in Fig.~\ref{fig3}. According to this plot, when the horizon area of the black hole increases, it leads $S_N$ and $S_{Sch}$ are also increased. Moreover, for fixed values of horizon area of black hole, it results that $S_{N}<S_{Sch}$, and the bigger $A$, the smaller difference is.

Considering $A=16\pi G^2 M^2$, the entropy formulation in equation \eqref{NonLocEnt} can be re-expressed as

\begin{equation}\label{NLE_Mod}
S_N=\frac{1}{4G}\left[48\sqrt{\pi}GMl e^{-\frac{G^2 M^2}{l^2}}+8\pi\left(2G^2 M^2-3l^2\right)~\text{erf}\left(\frac{GM}{l}\right)\right]
\end{equation}

The associated horizon temperature reads in this case as

\begin{equation}\label{NLE_Temp}
 \frac{1}{T}=8\pi GM\left[\text{erf}\left(\frac{GM}{l}\right)-\frac{2GM}{\sqrt{\pi} l}e^{-\frac{G^2 M^2}{l^2}}\right]
\end{equation}

Using \eqref{NLE_Mod} and \eqref{NLE_Temp} in the definition of the number of DoF $N=4S$, the analogous form to the standard equipartition law is generated as

\begin{equation}
M=\frac{1}{2}\frac{\left[\text{erf}\left(\frac{GM}{l}\right)-\frac{2GM}{\sqrt{\pi} l}e^{-\frac{G^2 M^2}{l^2}}\right]}{\left[\frac{3l}{\sqrt{\pi}GM} e^{-\frac{G^2 M^2}{l^2}}+\left(1-\frac{3l^2}{2G^2 M^2}\right)~\text{erf}\left(\frac{GM}{l}\right)\right]}NT
\end{equation}

Clearly, a large $A$ is equivalent to a large $M$ and in that case, the standard equipartition law can easily be realized from the above expression.\\

\begin{figure}
\centering
\includegraphics[width=11cm]{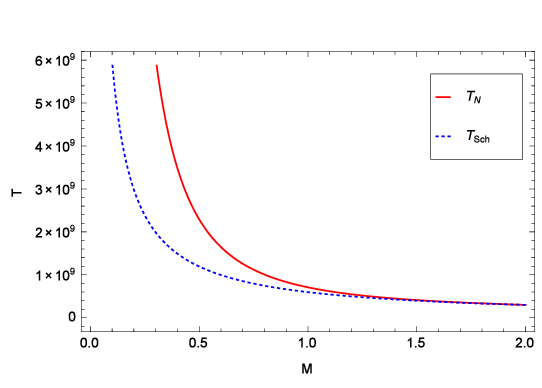}
\caption{\label{fig4}Comparison between modified black hole temperature due to Non-local gravity, $T_N$, and the Schwarzschild black hole temperature, $T_{Sch}$, versus
black hole mass $M$ for $l=10^{-6}$.}
\end{figure}

Next, in Fig.~\ref{fig4}, we plotted the modified black hole temperature due to Non-local gravity, $T_N$ (Eq.~\eqref{NLE_Temp}, and the Schwarzschild black hole temperature, $T_{Sch}$, versus black hole mass $M$. As it is shown, for both models, when mass increases, the temperature  of black hole decreases. Also, for fixed mass it concluded that $T_{N}>T_{Sch}$, which the difference between temperatures decrease with increasing mass.

Following equation \eqref{heat}, the expression of the heat capacity for this model takes the form

\begin{equation}\label{Nonl}
C=-8\pi G M^2 \frac{\left[\text{erf}\left(\frac{GM}{l}\right)-\frac{2GM}{\sqrt{\pi} l}e^{-\frac{G^2 M^2}{l^2}}\right]^2}{\left[\text{erf}\left(\frac{GM}{l}\right)-\frac{2GM}{\sqrt{\pi}l}e^{-\frac{G^2 M^2}{l^2}}\left(1-\frac{2G^2 M^2}{l^2}\right)\right]}
\end{equation}

The criteria for the thermodynamic stability therefore is

\begin{equation}
\text{erf}\left(\frac{GM}{l}\right)<\frac{2GM}{\sqrt{\pi}l}e^{-\frac{G^2 M^2}{l^2}}\left(1-\frac{2G^2 M^2}{l^2}\right)
\end{equation}

However, for considerably large $M$ compared to the length scale, the value of the Gauss error function is close to unity. Subsequently, the above inequality reduce to $1<0$ for $l\to 0$ and therefore our result can be confirmed with the expected notion of the unstable nature of the black hole in the classical format.

\begin{figure}
\centering
\includegraphics[width=11cm]{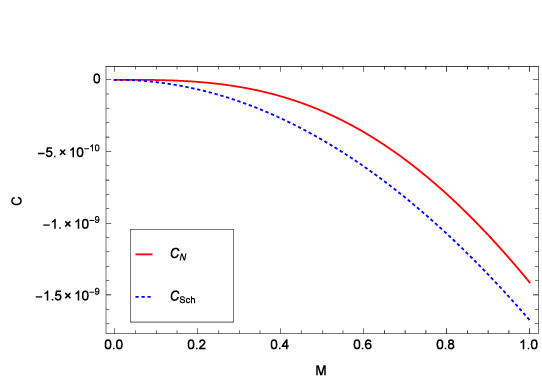}
\caption{\label{fig8}Comparison between the heat capacity of Non-local gravity, $C_N$, and the Schwarzschild heat capacity, $C_{Sch}$, versus
black hole mass $M$ for $l=10^{-6}$.}
\end{figure}

In figure (\ref{fig8}), we depict the heat capacities of non-local gravity, denoted as $C_N$ (Eq.~\eqref{Nonl}), and the Schwarzschild heat capacity, $C_{Sch}$, as functions of mass $M$. As anticipated, both heat capacities exhibit negativity. Specifically, as the mass $M$ increases, both $C_N$ and $C_{Sch}$ decrease, albeit at distinct rates. Notably, for a fixed mass, we observe that  $C_{Sch}$ is smaller than $C_N$.

\subsection{Gradient Field Gravity Correction}

The third alteration from the Newtonian potential is resulted by assuming the gradient field gravity. The potential function in this case is described as \cite{Lazar}

\begin{equation}
\phi=-\frac{GM}{R}\left[1-\frac{1}{{a_1}^2-{a_2}^2}\left({a_1}^2 e^{-\frac{R}{a_1}}-{a_2}^2 e^{-\frac{R}{a_2}}\right)\right]
\end{equation}

Here, the potential function is a resultant of a long-range Newtonian potential and a short-range bi-Yukawa-type potential \cite{Kamel}. The two parameters $a_1$ and $a_2$ are known as the internal characteristic length scale parameters. This potential implies a potentially measurable correction to gravity in the near field \cite{moi1,moi2}. However, the original Newtonian potential is recovered in the far field .\\

Due to the small-scale correction to the standard potential function, the modified form of the entropy becomes

\begin{equation}
S_G=\frac{1}{4G}\left[A+\frac{1}{{a_1}^2-{a_2}^2}\left\{{a_1}^2\left[4{a_1}^2\left(\sqrt{\frac{\pi A}{{a_1}^2}}+2\pi\right)e^{-\sqrt{\frac{A}{4\pi {a_1}^2}}}\right]-{a_2}^2\left[4{a_2}^2\left(\sqrt{\frac{\pi A}{{a_2}^2}}+2\pi\right)e^{-\sqrt{\frac{A}{4\pi {a_2}^2}}}\right]\right\}\right]
\end{equation}

Note that similar to the discussion we mentioned for the case of non-local gravity, so far, no experimental bounds have been obtained for the values of two parameters $a_1$ and $a_2$ \cite{moi1}.

\begin{figure}
\centering
\includegraphics[width=11cm]{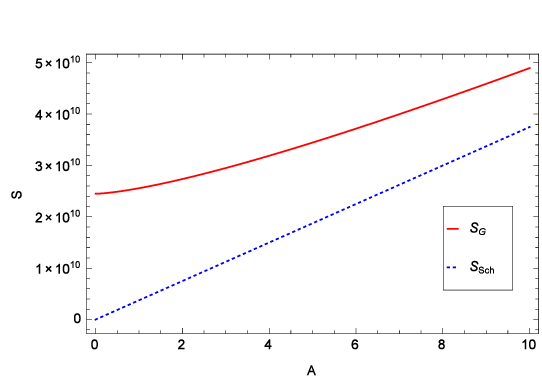}
\caption{\label{fig5}Comparison between Gradient field gravity entropy, $S_G$, and Schwarzschild entropy, $S_{Sch}$,
versus area $A$, for $a_1=10^{-9}$ and $a_2=10^{-10}$.}
\end{figure}

Now, in Fig.~\ref{fig5}, we have graphically depicted the modified entropy due to Gradient field gravity, ($S_G$), and the Schwarzschild entropy, ($S_{Sch}$), as functions of the black hole’s horizon area $A$. Our analysis reveals that as the horizon area of the black hole increases, both $S_G$ and $S_{Sch}$ also increase but with different rates. Furthermore, for fixed values of $A$, we observe that $S_G$ is bigger than $S_{Sch}$, however this difference becomes smaller as $A$ grows larger.

Using the definition $A=16\pi G^2 M^2$, the above entropy function can be alternatively expressed as

\begin{equation}\label{GFG_Ent}
S_G=\frac{1}{4G}\left[16\pi G^2 M^2+\frac{8\pi}{{a_1}^2-{a_2}^2} \left\{{a_1}^4 e^{-\frac{2GM}{a_1}}\left(1+\frac{2GM}{a_1}\right)-{a_2}^4 e^{-\frac{2GM}{a_2}}\left(1+\frac{2GM}{a_2}\right)\right\}\right]
\end{equation}

The associated temperature of the horizon will be

\begin{equation}\label{GFG_Temp}
 \frac{1}{T}=8\pi GM\left[1+\frac{1}{{a_1}^2-{a_2}^2}\left({a_2}^2 e^{-\frac{2GM}{a_2}}-{a_1}^2 e^{-\frac{2GM}{a_1}}\right)\right]
\end{equation}

Using \eqref{GFG_Ent} and \eqref{GFG_Temp} into the definition of the number of DoF $N=4S$, the revised form of the classical equipartition law in this case will be

\begin{equation}
 M=\frac{1}{2}\frac{\left[1+\frac{1}{{a_1}^2-{a_2}^2}\left({a_2}^2 e^{-\frac{2GM}{a_2}}-{a_1}^2 e^{-\frac{2GM}{a_1}}\right)\right]}{\left[1+\frac{1}{4G^2 M^2 ({a_1}^2-{a_2}^2)}\left\{{a_1}^4 e^{-\frac{2GM}{a_1}}\left(1+\frac{2GM}{a_1}\right)-{a_2}^4 e^{-\frac{2GM}{a_2}}\left(1+\frac{2GM}{a_2}\right)\right\}\right]}NT
\end{equation}

\begin{figure}
\centering
\includegraphics[width=11cm]{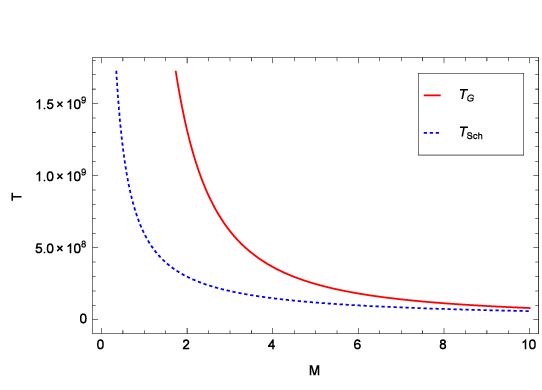}
\caption{\label{fig6}Comparison between Gradient field gravity temperature, $T_Y$, and the Schwarzschild black hole temperature, $T_{Sch}$, versus
black hole mass $M$ for $a_1=10^{-9}$ and $a_2=10^{-10}$.}
\end{figure}

From the above expression, it can be noticed that the standard form of the equipartition law can be achieved for far field.\\
Now, to show our results more clear, we compare the modified black hole temperature due to Gradient field gravity, $T_G$ (Eq.~\eqref{GFG_Temp}) with the Schwarzschild black hole temperature, $T_{Sch}$, as functions of the black hole mass $M$, in Fig.~\ref{fig6}. Evidently, both models exhibit a decrease in temperature as mass increases, although their slopes differ. Additionally, for a fixed mass, we find that $T_{G}>T_{Sch}$, with the temperature difference diminishing as mass increases.

The expression for the heat capacity in this model follows from the equation \eqref{heat}

\begin{equation}\label{CGradiant}
C=-8\pi GM^2\frac{\left[1+\frac{1}{{a_1}^2-{a_2}^2}\left({a_2}^2 e^{-\frac{2GM}{a_2}}-{a_1}^2 e^{-\frac{2GM}{a_1}}\right)\right]^2}{1+\frac{1}{{a_1}^2-{a_2}^2}\left[{a_2}^2\left(1-\frac{2GM}{a_2}\right)e^{-\frac{2GM}{a_2}}-{a_1}^2\left(1-\frac{2GM}{a_1}\right)e^{-\frac{2GM}{a_1}}\right]}
\end{equation}

Consequently, for a thermal stability, one must have

\begin{equation}
1+\frac{1}{{a_1}^2-{a_2}^2}\left[{a_2}^2\left(1-\frac{2GM}{a_2}\right)e^{-\frac{2GM}{a_2}}-{a_1}^2\left(1-\frac{2GM}{a_1}\right)e^{-\frac{2GM}{a_1}}\right]<0
\end{equation}

The above relation is coherent to the classical result in far fields, where black holes are considered to be thermally unstable in nature. This can be easily checked from the above inequality that when $a_1 \to 0$ and $a_2 \to 0$, we simply achieve $1<0$ indicating black hole instability in classical format.

\begin{figure}
\centering
\includegraphics[width=11cm]{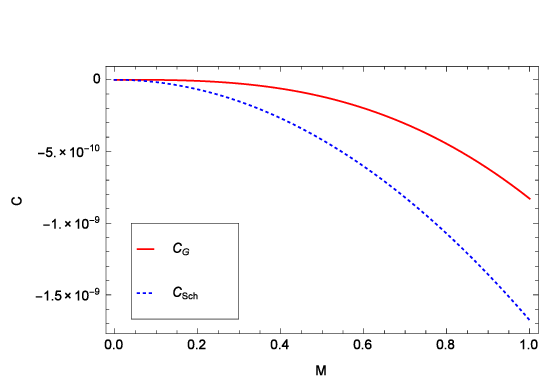}
\caption{\label{fig9}Comparison between the heat capacity of Gradient field gravity, $C_G$, and the Schwarzschild heat capacity, $C_{Sch}$, versus
black hole mass $M$ for $a_1=10^{-9}$ and $a_2=10^{-10}$.}
\end{figure}

Finally, in figure (\ref{fig9}), we visualize the heat capacities of Gradient field gravity, $C_G$ (from Eq.~\eqref{CGradiant}), and the Schwarzschild heat capacity, $C_{Sch}$, in terms of mass $M$. According to the plot, if $M_1>M_2$, both $C_G$ and $C_{Sch}$ decrease, although the decreasing slope for $C_{Sch}$ is steeper than $C_G$. Also, it is concluded that for a fixed mass, $C_{Sch}<C_G$.

\section{Conclusion}
In this paper, we investigated the thermodynamic behavior of black holes with a class of corrected entropies \cite{Kamel}. These entropies are attached to several modified gravity proposals, including Yukawa Potential Correction, Non-local Gravity Correction, and Gradient Field Gravity Correction, which were introduced to regularize singularities at small scales within the framework of entropic gravity. For each proposal, we analyzed the stability of the thermodynamic system by determining the black hole horizon temperature from the modified entropy. We then applied the definition of the number of degrees of freedom of the horizon to obtain the modified equipartition law, and finally, we used the heat capacity to assess the condition for stable thermodynamics. Through this analysis, we aimed to gain insights into the thermodynamic behavior of these modified gravity proposals and evaluate their potential for providing a consistent and coherent description of gravitational phenomena.
We would like to mention here that the stability of the end state of the black hole evaporation process has some ambiguity and several physicists believe that the final stage of the black hole evaporation might lead to a stable remnant when prescribed in a non-perturbative theory of quantum gravity \cite{rovelli,lorenzo, christo}. In this alternative theory framework, it has been argued that the semi-classical approaches in deriving the Hawking radiation become inconsistent, most possible near the Planck scale $l_p$. The stability of such a stage is widely investigated in different gravity theories \cite{maluf,rovelli2,Konoplya,Mukhanov,Kuntz,khan}.\\
Our analysis of the heat capacity for the Yukawa Potential Correction model revealed that the stability of black holes cannot be achieved in the far field when the correction parameter is very small. This result aligns with the expectation that black holes are inherently thermally unstable. Similarly, for the Non-local Gravity Correction model, the heat capacity was expressed in terms of Gauss error functions, and we found that for large black hole masses, the value of the function approaches unity, consistent with the classical notion of black hole thermal instability. Finally, for the Gradient Field Gravity Correction model, we obtained a condition for thermal stability that is consistent with the classical result in the far field, where black holes are expected to be thermally unstable. Moreover, in figs. (\ref{fig7}), (\ref{fig8}) and (\ref{fig9}) we plotted Yukawa heat capacity, Non-local Gravity heat capacity and also Gradient field gravity capacity versus mass respectively which concluded that negative heat capacities for all three models as expected. Note that, since the heat capacity of a black hole can be either positive or negative due to the black hole’s size (See \cite{Angsachon,Promsiri,Li} and some references therein), and in our cases, all heat capacities are obtained as negative values, it can be interpreted that these black holes are thermally unstable and tend to radiate energy, leading to a phase transition. These findings provide insights into the thermodynamic behavior of these modified gravity models and their relation to the classical understanding of black hole stability.


\begin{thebibliography}{99}

\bibitem{Bekenstein} J. D. Bekenstein, Phys. Rev. D 7, 2333 (1973).
\bibitem{Hawking} S. W. Hawking, Commun. Math. Phys. 43, 199 (1975).
\bibitem{Jacobson} T. Jacobson, Phys. Rev. Lett. 75, 1260 (1995).
\bibitem{Padmanabhan1} T. Padmanabhan, Phys. Rept. 406, 49 (2005).
\bibitem{Padmanabhan2} T. Padmanabhan, J. Phys. Conf. Ser. 306, 012001 (2011).
\bibitem{Elizalde} E. Elizalde and P. J. Silva, Phys. Rev. D 78, 061501 (2008).
\bibitem{Chirco} G. Chirco and S. Liberati, Phys. Rev. D 81, 024016 (2010).
\bibitem{Sakharov} A. D. Sakharov, Sov. Phys. Dokl. 12, 1040 (1968).
\bibitem{Visser} M. Visser, Mod. Phys. Lett. A 17, 977 (2002).
\bibitem{Bohm} D. Bohm, PHYSICS (GB), 3.2, pp. 139-168 (June 1973).
\bibitem{Talbot} M. Talbot, HarperCollins Publishers, Inc., New York (1991).
\bibitem{Susskind} L. Susskind, J. Lindesay , World Scientific, December (2004).
\bibitem{Cai} R. G. Cai, L. M. Cao, and N. Ohta, Phys. Rev. D 81, 084012 (2010).
\bibitem{Qiu} T. Qiu and E. N. Saridakis, Phys. Rev. D 85, 043504 (2012).
\bibitem{Komatsu} N. Komatsu and S. Kimura, Phys. Rev. D 90, 123516 (2014).
\bibitem{Shi} B. Shi, Phys. Rev. Research 2, 023132 (2020).
\bibitem{Verlinde} E. Verlinde, J. High Energy Phys. 2011, 29 (2011).
\bibitem{Sheykhi1} A. Sheykhi, Phys. Rev. D 81, 104011 (2010).
\bibitem{Meissner} K. A. Meissner, Class. Quantum Grav. 21, 5245 (2004).
\bibitem{Sheykhi2}  A. Sheykhi and S. H. Hendi, Phys. Rev. D 84, 044023 (2011).
\bibitem{Das} S. Das, S. Shankaranarayanan, and S. Sur, Phys. Rev. D 77, 064013 (2008).
\bibitem{Nicolini} P. Nicolini, Phys. Rev. D 82, 044030 (2010).
\bibitem{Merino} A. Mart´ınez-Merino, O. Obreg´on, and M. P. Ryan, Jr., Phys. Rev. D 95, 124031 (2017).
\bibitem{Obreg} O. Obreg´on, Entropy 12(9), 2067 (2010).
\bibitem{Moradpour} H. Moradpour, A. Sheykhi, C. Corda, and I. G. Salako, Phys. Lett. B 783, 82 (2018).
\bibitem{Gao} C. Gao, Phys. Rev. D 81, 087306 (2010).
\bibitem{Pazy} E. Pazy, Phys. Rev. D 87, 084063 (2013).
\bibitem{Abreu1} E. M. C. Abreu, J. A. Neto, E. M. Barboza, Jr., and R. C. Nunes, Int. J. Mod. Phys. A 32, 1750028 (2017).
\bibitem{Ourabah1} K. Ourabah, E. M. Barboza, Jr., E. M. C. Abreu, and J. A. Neto, Phys. Rev. D 100, 103516 (2019).
\bibitem {Nozari1} K. Nozari, P. Pedram,  M. Molkara, Int. J. Theor. Phys. 51, 1268 (2012).
\bibitem{Nozari2} K. Nozari, S. Akhshabi,  Phys. Lett. B 700, 91 (2011).
\bibitem{Kaul} R. K Kaul and P. Majumdar, Phys. Rev. Lett. 84, 5255 (2000).
\bibitem{Adler} R. J. Adler, P. Chen and D. I Santiago, Gen. Rel. Grav. 23, 2101 (2001).
\bibitem{Das1} S. Das, P. Majumdar and R. K. Bhaduri, Class. Quant. Grav. 19, 2355 (2002).
\bibitem{Myung} Y. S. Myung, Phys. Lett. B 579, 205 (2004).
\bibitem{Doma} M. Domagala and J. Lewandowski, Class. Quant. Grav. 21, 5233 (2004).
\bibitem{Chatterjee} A. Chatterjee and P. Majumdar, Phys. Rev. Lett. 92, Article ID 141301 (2004).
\bibitem{Ghosh} A. Ghosh and P. Mitra, Phys. Lett. B 616, 114 (2005).
\bibitem{Camelia} G. A. Camelia, M. Arzano, Y. Ling and G. Mandanici, Class. Quant. Grav. 23, 2585 (2006).
\bibitem{Nozari} K. Nozari and A. S. Sefiedgar, Gen. Rel. Grav. 39, 501 (2007).
\bibitem{Barun} B. Majumdar, Phys. Lett. B 703, 402 (2011).
\bibitem{Barun2} B. Majumdar, Gen. Rel. Grav. 45, 2403 (2013).
\bibitem{Kamel} K. Ourabah, Class. Quantum Grav., 41, 015010 (2024).
\bibitem{plus1} A. de Almeida, L. Amendola, and V. Niro,{J. Cosmol. Astropart. Phys. {08}, 012 (2018).}
\bibitem{plus2} A. Bessiri, K. Ourabah, and T. H. Zerguini, Phys. Scr. 96 125208 (2021).
\bibitem{plus3} I. De Martino, R. Lazkoz, and M. De Laurentis, {Phys. Rev. D {97}, 104067 (2018).}
\bibitem{plus4} J. T. Mendonça, Symmetry 13(6), 1007 (221).
\bibitem{Maneff1} G. Maneff, Comptes Rendus Acad. Sci. Paris 178, 2159 (1924).
\bibitem{Maneff2} G. Maneff, Comptes Rendus Acad. Sci. Paris 190, 963 (1930).
\bibitem{Maneff3} G. Maneff, Comptes Rendus Acad. Sci. Paris 190, 1374 (1930).
\bibitem{Hagihara} Y. Hagihara, Celestial Mechanics (MIT Press 1972).
\bibitem{Tohline} J. E. Tohline, The Internal Kinematics and Dynamics of Galaxies, IAU Symp. 10; E. Athanassoula, Ed.; Reidel: Dordrecht, The Netherlands, 1983; pp. 205–206.
\bibitem{Kuhn} J. R. Kuhn and L. Kruglyak, Astrophys. J. 313, 1 (1987).
\bibitem{Acedo} L. Acedo, Galaxies 5(4), 74 (2017).
\bibitem{Finzi} A. Finzi and F. A. E. Pirani, Mon. Not. R. Astron. Soc. 127, 21 (1963).
\bibitem{Giusti} A. Giusti, Phys. Rev. D 101, 124029 (2020).
\bibitem{Hoyle} C. D. Hoyle et al., Phys. Rev. D 70, 042004 (2004).
\bibitem{Smullin} S. J. Smullin et al., Phys. Rev. D 72, 122001 (2005).
\bibitem{Modesto1} L. Modesto, Phys. Rev. D 86, 044005 (2012).
\bibitem{Biswas} T. Biswas, E. Gerwick, T. Koivisto, and A. Mazumdar, Phys. Rev. Lett. 108, 031101 (2012).
\bibitem{Modesto} L. Modesto, T. de Paula Netto, and I. L. Shapiro, J. High Energy Phys. 04, 098 (2015).
\bibitem{Lazar} M. Lazar, Phys. Rev. D 102, 096002 (2020).
\bibitem{Abreu} E. M. C. Abreu and J. A. Neto, Eur. Phys. J. C, 80, 776, (2020).
\bibitem{Paddy} T. Padmanabhan, Class. Quantum Gravity 21, 4485 (2004).
\bibitem{Kom}  N. Komatsu, Eur. Phys. J. C 77, 229 (2017).
\bibitem{eqt1} E. M. C. Abreu et al., EPL 130, 40005 (2020).
\bibitem{eqt2} K. Ourabah
Phys. Rev. D 102, 043017 (2020).
\bibitem{eqt3} J. A. Neto, Physica A 391, 4320 (2012).
\bibitem{nonlocal1} L. Modesto, {Phys.
Rev. D {86}, 044005 (2012).}
\bibitem{nonlocal2} T. Biswas, E. Gerwick, T. Koivisto, and A. Mazumdar,
{Phys. Rev. Lett. {108}, 031101 (2012).}
\bibitem{nonlocal3} L. Modesto, T. de Paula Netto, and I. L. Shapiro, {J. High Energy Phys. {04}, 098 (2015).}
\bibitem{Lazar} M. Lazar, {Phys. Rev. D {102}, 096002 (2020).}
\bibitem{Kapner} D. J. Kapner et al., Phys. Rev. Lett. 98, 021101 (2007).
\bibitem{Lee} J. G. Lee et al., Phys. Rev. Lett. 124, 101101 (2020).
\bibitem{Martino} I. De Martino, R. Lazkoz, and M. De Laurentis, Phys. Rev. D 97, 104067 (2018).
\bibitem{moi1} K. Ourabah, {Sci. Rep. {12}, 15717 (2022).}
\bibitem{moi2} K. Ourabah, Eur. Phys. J. Plus 138, 55 (2023).
\bibitem{rovelli} C. Rovelli and F. Vidotto, Int. J. Mod. Phys. D, 23, 1442026 (2014).
\bibitem{lorenzo} T. D. Lorenzo and A. Perez, Phys. Rev. D, 1, 124018 (2016).
\bibitem{christo} M. Christodoulou and F. D'Ambrosio, arXiv: 1801.03027, (2018).
\bibitem{maluf} R. V. Maluf and J. C. S. Neves, Phys. Rev. D, 97, 104015 (2018).
\bibitem{rovelli2} C. Rovelli and F. Vidotto, Universe, 2018 4(11), 127.
\bibitem{Konoplya} R. A. Konoplya and A. F. Zinhailo, Phys. Rev. D, 99, 104060 (2019).
\bibitem{Mukhanov} A. H. Chamseddine, V. Mukhanov and T. B. Russ, J. High Energy Phys., 104 (2019).
\bibitem{Kuntz} I. Kuntz and R. de Rocha, Theor. Phys., 80, 478 (2020).
\bibitem{khan} Y. H. Khan, S. Upadhyay and P. A. Ganai, Mod. Phys. Lett. A, 36, 2130023 (2021).
\bibitem {Angsachon} T. Angsachon, K. Ruenearom, Theor. Math. Phys. 217, 1 (2023).
\bibitem {Promsiri} C. Promsiri , E.  Hirunsirisawat , W. Liewrian , Phys. Rev. D 104, 064004 (2021).
\bibitem{Li} H. L. Li, W. Li, Int. J. of The. Phy.  59 , 3032 (2020).
\end{thebibliography}
\end{document}